\documentclass[12pt]{article} 
\usepackage[american]{babel}
\usepackage{epsfig}
\usepackage{cite}
\usepackage{hyperref}
\newcommand{\spmine}{1.05}
\newcommand{\myspace}{\edef\baselinestretch{\spmine}\Large\normalsize}

\makeatletter
\def\@eqnnum{{\normalfont \normalcolor [\theequation]}}
\makeatother

\newcommand{\be}{\begin{equation}}
\newcommand{\ee}{\end{equation}}
\newcommand{\ba}{\begin{eqnarray}}
\newcommand{\ea}{\end{eqnarray}}

\newcommand{\ben}{\begin{displaymath}}
\newcommand{\een}{\end{displaymath}}
\newcommand{\ban}{\begin{eqnarray*}}
\newcommand{\ean}{\end{eqnarray*}}
\newcommand{\brr}{\begin{array}}
\newcommand{\err}{\end{array}}

\renewcommand{\theequation}%
  {\thesection.\arabic{equation}}
\newcommand{\bc}{\begin{center}}
\newcommand{\ec}{\end{center}}

\begin{document}

\title{THERMAL EFFECTS IN  PHOTOEMISSION
 AND ELECTRON-PHONON COUPLINGS OF FULLERENE}

\author{Andrea Bordoni and Nicola Manini \\
\it
Dipartimento di Fisica, Universit\`a di Milano,\\
\it
Via Celoria 16, 20133 Milano, Italy\\
\it and INFM, Unit\`a di Milano, Milano, Italy
}


\date{}
\maketitle
\begin{abstract}
We show that thermal effects play a relevant role in the determination of
the electron-phonon couplings based on the intensity of the phonon-shakeup
peaks in photoemission spectra.
In particular, we re-consider the determination of the electron-phonon
couplings of fullerene based on a fit of the photoemission spectrum of
C$_{60}^-$ in gas phase \cite{Gunnarsson}.
We show that taking thermal effects into account reduces the obtained
couplings by approximately 10\%.
\end{abstract}
%
%

\thispagestyle{empty}		  
\myspace

\section{\centerline{INTRODUCTION}}

Many of the interesting phenomena observed in electron-donor doped
fullerides are related to electron-phonon coupling of the partly filled
band derived from the lowest unoccupied molecular orbital (LUMO) of
C$_{60}$.
These couplings, in turn, are mainly of molecular origin, involving the
optical phonons derived from the molecular modes of $A_g$ and $H_g$
symmetry.
Several different determinations of the electron-phonon couplings of the
LUMO are available on the market, both from calculations
\cite{vzr,Schl,Antropov,Manini01}, and derived from experimental data
\cite{Gunnarsson}.
All calculations based on density-functional theory (DFT) invariably turn out
much smaller couplings than those obtained through fitting the experimental
photoemission spectrum (PES) of gas-phase C$_{60}^-$ (see for example
Table~\ref{Tabella_Modi}).
This spectrum is affected by the electron-phonon shakeup associated to 
relaxation after the sudden ejection of the extra electron from the LUMO.
That fit was realized on the basis of a zero-temperature model for the
electron-phonon coupled system.
The authors of Ref.~\citen{Gunnarsson} estimate the vibrational temperature
of the sample ions of the order of 200~K.
We claim that even such a low temperature could affect significantly the
spectrum, in a way not unlike an increase of the couplings.
We conclude that a determination of the couplings {\it via} a model
including realistic thermal effects must turn out smaller  couplings than a
zero-temperature model.

\begin{table}
\begin{center}
\begin{tabular}{lccll}
\hline
\hline
Mode$^{(j)}$ & Exp.\ freq.\ (cm$^{-1}$) & DFT Energy (meV)
& $g_{\Lambda_j}$ PES\cite{Gunnarsson} & $g_{\Lambda_j}$ DFT\cite{Manini01} \\
\hline
$A_g^{(1)}$ &  496 & & 0.0 & 0.157 \\
$A_g^{(2)}$ &  1470& & 0.851 & 0.340 \\
\hline
$H_g^{(1)}$ &  271 & 32  & 0.824 & 0.412 \\
$H_g^{(2)}$ &  437 & 53 & 0.941 & 0.489 \\
$H_g^{(3)}$ &  710 & 89 & 0.421 & 0.350 \\
$H_g^{(4)}$ &  774 & 97 & 0.474 & 0.224 \\
$H_g^{(5)}$ &  1099& 139 & 0.325 & 0.193 \\
$H_g^{(6)}$ &  1250& 158 & 0.197 & 0.138 \\
$H_g^{(7)}$ &  1428& 180 & 0.339 & 0.315 \\
$H_g^{(8)}$ &  1575& 197 & 0.376 & 0.289 \\
\hline
\hline
\end{tabular}
\end{center}
\caption{\label{Tabella_Modi}
Experimental frequencies of the vibrational modes of fullerene linearly
coupled to the $t_{1u}$ LUMO.
Computed energies and dimensionless coupling constants.
}
\end{table}

\section{\centerline{THE MODEL}}\label{Model}

The standard formulation of the linear electron-phonon model
\cite{ob69,AMT,Gunnarsson} describing the JT coupling of the orbitally
degenerate $t_{1u}$ molecular triplet with the vibrations of symmetry
$\Lambda = A_g,\,H_g$ is the following:
\begin{eqnarray}
\label{modelhamiltonian}
\hat{H} &=& \hat{H}_0 + \hat{H}_{\rm vib} +
\hat{H}_{\rm e-v} \,,\\
\hat{H}_0     &=& \epsilon \, \sum_{m \sigma}
\hat{c}_{m \sigma}^\dagger \hat{c}_{m \sigma} \,, \\
\label{vib-hamiltonian}
\hat{H}_{\rm vib} &=& \frac 12  \sum_{j \mu} \hbar \omega_{\Lambda j}
\left(\hat{P}_{j\mu}^2+\hat{Q}_{j\mu}^2\right) \,, \\
\label{JT-hamiltonian}
\hat{H}_{\rm e-v} &=& 
\, k^\Lambda
\sum_{{\sigma \, \mu m m'}}
C^{\Lambda \mu}_{m \; -m'} \, \sum_j
g_{\Lambda j} \hbar \omega_{\Lambda j}
\hat{Q}_{j\mu} \,
\hat{c}^\dagger_{m\sigma} \hat{c}_{m' \sigma } \,.
\end{eqnarray}
Here, $\hat{Q}_{j\mu} =\left(\hat{b}_{j\mu}^\dagger +
\hat{b}_{j\mu}\right)/\sqrt 2$ is the dimensionless normal-mode vibrational
coordinate (in units of the natural length scale of the harmonic
oscillator), and $\hat{P}_{j\mu}$ the corresponding conjugate momentum for
mode $j$ of symmetry $\Lambda$ (either $A_g$: 2 modes, or $H_g$: 8
modes).
$\hat{c}^\dagger_{m\sigma}$ is a fermion operator creating a spin-$\sigma$
electron in orbital $m=0,\pm 1$ of the LUMO.
$m$, $m'$ and $\mu$ label components within the degenerate multiplets, for
example according to the $C_5$ character in the ${\cal I}\supset D_5\supset
C_5$ group chain \cite{hbyh,Butler81}.
In particular, $\mu=0,\pm1,\pm2$ for $H_g$ modes and $\mu=0$ only for $A_g$
modes.
$C^{\Lambda \mu}_{m m'}$ are Clebsch-Gordan coefficients \cite{Butler81} of
the icosahedral group ${\cal I}$ that couple two $t_{1u}$ tensor operators
and a $\Lambda$ tensor operator to a global scalar $A$ operator (in
practice these coefficients equal spherical Clebsch-Gordan coefficients for
coupling two $l=1$ angular momenta to $L=0$ for $\Lambda=A_g$ and $L=2$ for
$\Lambda=H_g$).
Numerical factors $k^{A}=\sqrt{3}$, $k^{H}=\sqrt{3/2}$ (which could
otherwise be re-absorbed into the definition of $g_{\Lambda j}$) are included
to make contact with previous notation \cite{AMT,Manini01}.
$g_{\Lambda j}$ are precisely the electron-phonon coupling parameters whose
determination is addressed in this work.

The $H_g$-mode coupling in the model [\ref{modelhamiltonian}] is of the
Jahn-Teller (JT) type: exact solution are known only for very large or very
small couplings.
The couplings of fullerene appear to be in an intermediate regime, for
which only numerical solutions are available.
Indeed, the authors of Ref.~\citen{Gunnarsson}, applied Lanczos
diagonalization to the Hamiltonian [\ref{modelhamiltonian}] on a truncated
harmonic basis, to obtain its ground state.
The analysis of this ground state in terms of the harmonic ladder of
oscillators of the non-JT neutral C$_{60}$ provided the intensities to
reconstruct the spectrum.
At any finite temperature one should in principle repeat the analysis not
only for the JT ground state, but also for a number of low-lying
excitations and sum the resulting spectra, each multiplied by the
corresponding Boltzmann weight.
An attempt to follow this route at any finite temperature such that $k_B T$
is comparable to even the lowest quanta of vibration ($\approx 30$~meV) is
beyond the possibilities of nowadays computers, due to the large number of
oscillators, and the associated rapidly increasing density of coupled
states: a reliable evaluation of the spectrum would require the accurate
determination of the wavefunction of several hundred thousand of the lowest
states in a Hilbert space of several million.

\begin{figure}
\centerline{
\epsfig{file=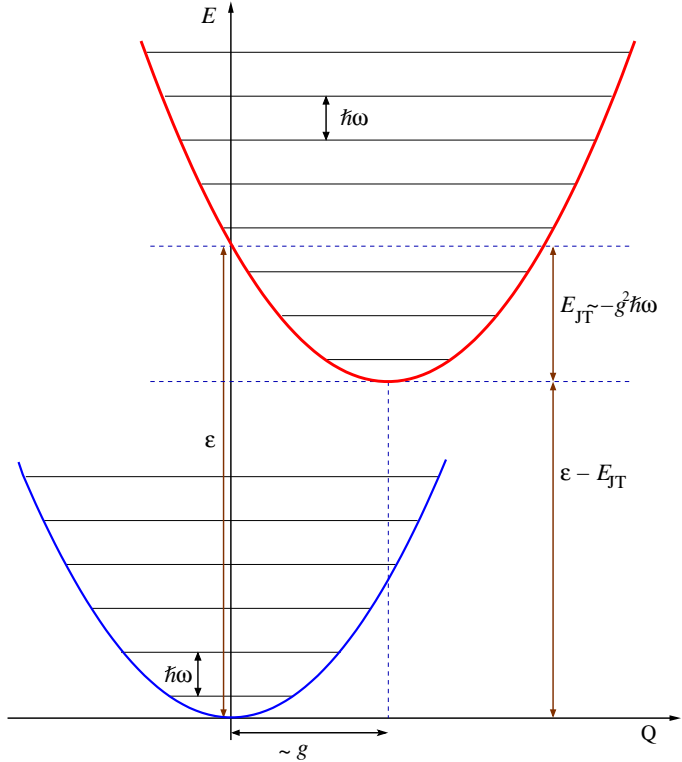,width=9cm,clip=}
}
\caption{\label{ev_scheme:fig}
The schematic ``Frank-Condon'' structure of the PES experiment, as seen
from the point of view of any given normal-mode coordinate $Q$.
Along each normal-mode direction, the displacement of the minimum of the
ionized configuration with respect to that of the neutral configuration is
proportional to the coupling of that mode.
}
\end{figure}

We propose here a more efficient method, based on a related much simpler
model, for which an analytical solution allows to carry out the calculation
even to fairly high temperature, with mild computational requirements.
The basic observation here is that in the experiment under examination the
final states (neutral C$_{60}$) involve simple harmonic ladders of phonons.
These are excited by an amount which is related to the displacement of the
adiabatic minimum of the initial C$_{60}^-$ with respect to that of the
final C$_{60}$, as illustrated in Fig.~\ref{ev_scheme:fig}.
The intricate details of the initial JT state should play only a minor role
in this spectrum, the main thing that needs to be made right being the
displacement between the two minima.
We replace therefore each of the $H_g$ modes (with the complication of the
full JT Hamiltonian [\ref{modelhamiltonian}]) with a set of five effective
$A_g$ modes, with the same frequency as the original fivefold-degenerate
mode, and a suitable coupling $g_{A}^{\rm eff}$.
It is important to introduce 5 degenerate oscillators per mode, to
reproduce correctly the increasing density of oscillator states with
energy, which is crucial to describe thermal effects.
The advantage of $A_g$ nondegenerate oscillators is that their wavefunction
expansion on a displaced basis is known analytically.
Note that this approximation could not possibly work in a context (such as
that of the PES from {\em neutral} C$_{60}$ \cite{Bruhwiler97,Manini03})
where the {\em final} states are affected by JT.

Briefly, starting from an initial state $|i\rangle=|v\rangle$ with $v$
quanta in the initial state, the required probability amplitude to end up
into a final state $|f\rangle=T^{-a}|w\rangle$ ($\hat{T}^{-a}$ is a space
translation operator and the distance $a = \frac{1}{2}\,g_A$) with $w$
quanta is
\begin{equation}
\label{a_mod_resid}
\langle f| \hat{c}_{m \sigma}^\dagger | i \rangle=
\langle \varphi_{w} \vert T^{a}\vert \varphi_{v}\rangle
= 
\sqrt{\frac{2^v\, v!}{2^w\, w!}}\; e^{-\frac14 a^2}\,a^{w-v}\,
L_{v}^{w-v}\!\left(\frac{a^{2}}{2}\right)
\end{equation}
where $L_{r}^{\lambda}(z)$ are the Laguerre polynomials of degree $r$:
$L_{r}^{\lambda}(z)\!=\!
\frac{z^{-\lambda}e^{z}}{r!}\frac{\mathrm{d}^{r}}{\mathrm{d}z^{r}}
\Big(z^{r+\lambda}e^{-z}\Big)$.
The spectral function is then computed according to Fermi's golden rule in
the sudden approximation:
\begin{equation}\label{FermiGR}
I_i(h\nu) = 
\frac{2\pi}{\hbar}
\sum_f|\langle f|V_I|i\rangle|^2\delta(h\nu- (E_f-E_i)),
\end{equation}
where $E_i$ and $E_f$ have trivial harmonic expressions.
Assuming a Boltzmann population of the initial states
\begin{equation}\label{Boltzmann}
 P(i)= \frac{\exp\left[-\frac{E_i}{k_{\rm B}T}\right] }
     { \sum_l\exp\left[-\frac{E_l}{k_{\rm B}T}\right] }\,,
\end{equation}
the final spectrum is obtained through the following thermal average:
\begin{equation}\label{thermalaverage}
 I(h\nu) = \sum_{i} P(i) \, I_{i}(h\nu)\,.
\end{equation}
In the problem of C$_{60}$ at hand, the total number of oscillators is
$5\times 8+2=42$.
To combine the spectral contributions of individual oscillators, the
collective matrix element in [\ref{FermiGR}] is the product of the matrix
elements of the individual oscillators, while the position $(E_f-E_i)$ of
the peak is of course the sum of the energy change of the individual
oscillators:
\begin{eqnarray}\label{combinepole}
\langle f|V_I|i\rangle \!\!\!&=&\!\!\!
\langle w_1,w_2,...|V_I|v_1,v_2\rangle =
\langle w_1|T^{a_1}|v_1\rangle 
\langle w_2|T^{a_2}|v_2\rangle ...\\\label{combineen}
E_f-E_i \!\!\!&=&\!\!\!
(E_{w_1}+E_{w_2}+...) -(E_{v_1} + E_{v_2}+...)=\\\nonumber
\!\!\!&=&\!\!\!(E_{w_1}-E_{v_1}) + (E_{w_2}-E_{v_2}) + ...
\end{eqnarray}
The delta-function peaks are finally broadened into Gaussians of 20~meV
HWHM to account for the final resolution of the spectrometer.
For more details on this simple model, the reader is referred to
Ref.~\citen{Gattari03}.

\begin{figure}
\centerline{
\epsfig{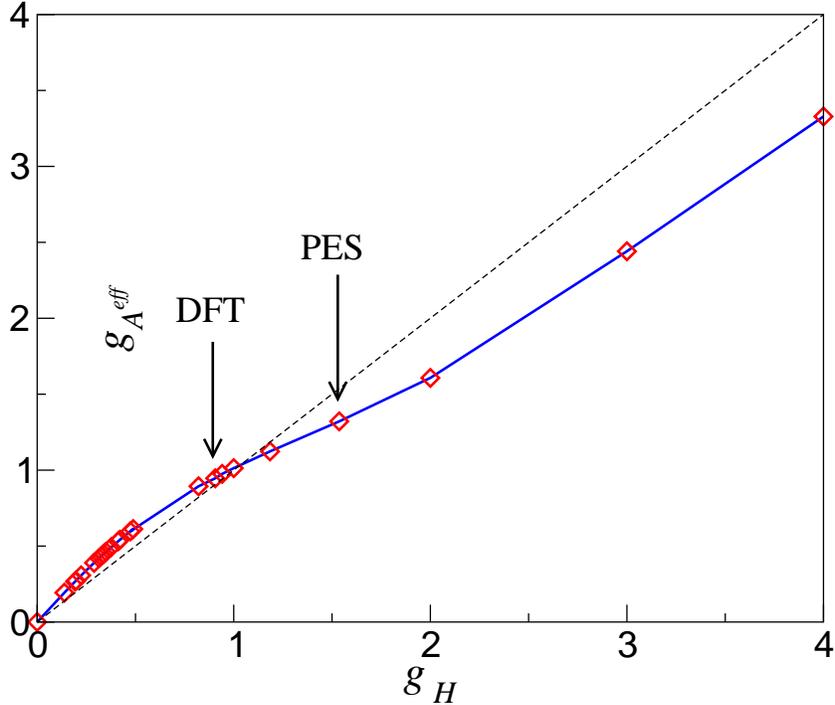}
}
\caption{\label{geff:fig}
The best fit value of $g_{A}^{\rm eff}$ as a function of  $g_H$, according
to a $T=0$ calculation.
Arrows point to the values of the total $\overline{g}_H$ according to the
DFT \cite{Manini01} and PES \cite{Gunnarsson} estimates.
}
\end{figure}

To determine the value of the effective couplings $g_{A_j}^{\rm eff}$, we
fit the $T=0$ spectrum of a single $H_g$ mode (computed by exact Lanczos
diagonalization of [\ref{modelhamiltonian}]) with that of 5 degenerate
$A_g$ oscillators of the same frequency.
By repeating this fit for a range of couplings, we obtain the curve
of Fig.~\ref{geff:fig}.
We have then verified that the temperature dependence of the spectrum based
on the zero-temperature fit is in very good accord to that obtained by
exact diagonalization (which, for this single-mode case, can be carried out
up to moderately high temperature).

\begin{figure}
\centerline{
\epsfig{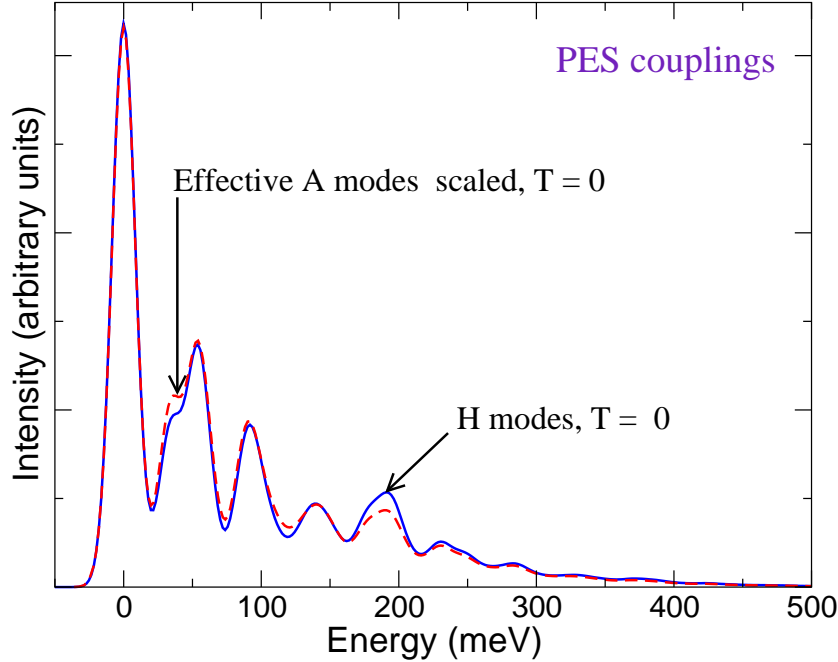}
}
\caption{\label{T_0.x_ModiH_A_eff_scal_orig:fig}
$T=0$ comparison between the exact JT Lanczos spectrum (solid line) and that
obtained by the simplified effective-$A$ modes model (dashed line).
Comparison done for PES couplings \cite{Gunnarsson}, Gaussian broadening of
10~meV HWHM, including only $H_g$ modes.
}
\end{figure}

For treating many modes, it is necessary to take into account the
``collective'' nature of the JT interaction.
Therefore, instead of converting each individual $g_{H j}$ through the
relation of Fig.~\ref{geff:fig}, we consider the ``total coupling''
\cite{ob72,ManyModes} $\overline{g}_H =\left( \sum_j
g_{Hj}^2\right)^{1/2}$, and convert $\overline{g}_H$ into a
$\overline{g}_A^{\rm eff}$, through Fig.~\ref{geff:fig}.
We then multiply each individual $g_{Hj}$ by the same ratio
$\overline{g}_A^{\rm eff}/\overline{g}_H$, to obtain the effective
$g_{Aj}^{\rm eff}$ couplings for each mode.
For the DFT coupling reported in Table~\ref{Tabella_Modi}, the ratio
$\overline{g}_A^{\rm eff}/\overline{g}_H = 1.045$, while for the PES
couplings $\overline{g}_A^{\rm eff}/\overline{g}_H = 0.859$.
On the basis of these couplings, we compute the $T=0$ spectrum, and in both
cases we obtain a good accord with the exact Lanczos calculation.
Figure~\ref{T_0.x_ModiH_A_eff_scal_orig:fig} illustrates this comparison
for the PES couplings.

\section{\centerline{RESULTS AND DISCUSSION}}

\begin{figure}
\centerline{
\epsfig{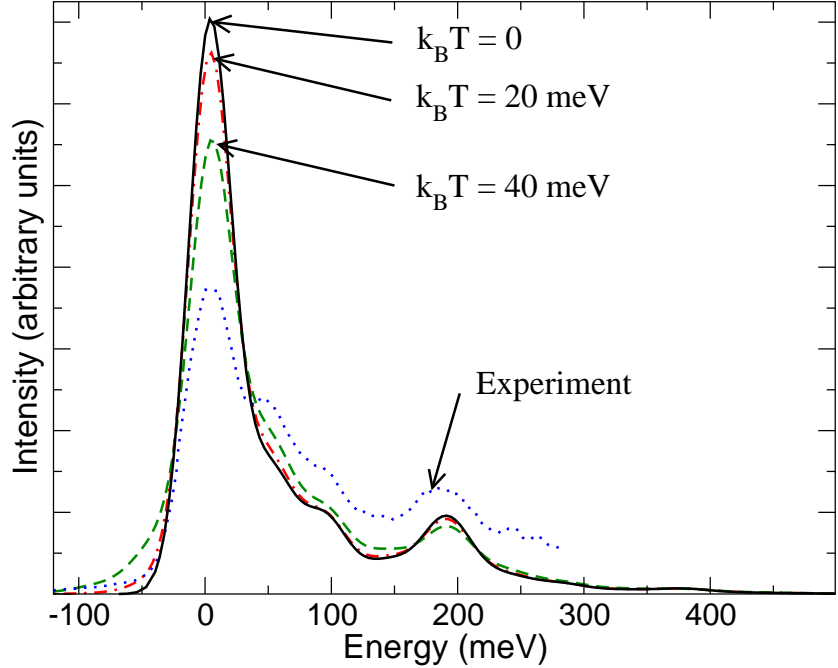}
}
\caption{\label{CExp_SoloA_T20_40_g20_cut:fig}
All-modes PES computed based on the effective-$A$ modes model at
$k_BT$=$20$ and $40$ meV, for the DFT couplings of Ref.~\citen{Manini01}.
The poor accord with experiment indicates that, even accounting for thermal
effects, the DFT couplings are indeed too small to account for the observed
transfer of spectral weight to the phonon satellites.
}
\end{figure}

We apply now the effective-$A$ modes model to the computation of the PES
at finite temperature.
Figure~\ref{CExp_SoloA_T20_40_g20_cut:fig} reports the thermal spectrum for
realistic values $k_BT = 20$, 40~meV, computed on the basis of the DFT
parameters \cite{Manini01}, rescaled as discussed above.
Clearly, even though thermal effects go in the direction of increasing the
transfer of weight to the phonon excitation and reducing the distance to
the experimental spectrum, the agreement remains poor, thus confirming that
the DFT parameters are indeed too small.
There is not a satisfactory explanation yet for this failure of DFT, which
is particularly surprising in light of the accurate values for the
couplings of the HOMO of C$_{60}$ obtained by the same method
\cite{Manini03}.

\begin{figure}
\centerline{
\epsfig{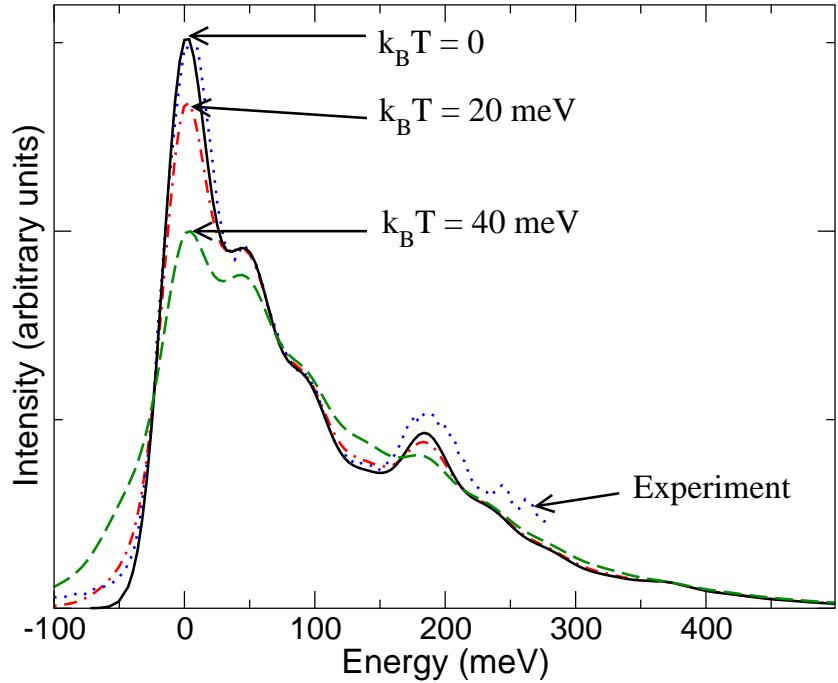}
}
\caption{\label{CExp_SoloA_T20_40_g20_Gunn_cut:fig}
All-modes PES computed based on the effective-$A$ modes model at
$k_BT$=$20$ and $40$ meV, for the couplings of Ref.~\citen{Gunnarsson},
obtained through a $T=0$ fit to the PES data, and compared to experiment.
}
\end{figure}

We then verify thermal effects on the spectrum based on the couplings
parameters fitted on the PES spectrum using the $T=0$ model
\cite{Gunnarsson}.
Figure~\ref{CExp_SoloA_T20_40_g20_Gunn_cut:fig} reports the thermal
spectrum for $k_BT$ = 0, 20, 40~meV.
Clearly, for $T=0$, the accord with experiment is good.
As $T$ is raised, more and more weight moves to the phonon satellites,
making the accord less good.
To restore a better accord at finite temperature, smaller couplings should
be considered, reduced by perhaps 10\%.
For a precise evaluation of the amount of this reduction, a reliable
estimate of the temperature of the sample is needed.
Unfortunately, the intensity of the ``pre-edge'' anti-Stokes feature is
only a rather indirect measure of temperature, since this intensity depends
crucially also on the value of the coupling of the involved low-frequency
phonon mode.
(This is clearly illustrated in the comparison of the $k_BT=40$~meV dashed
curves of Figs.~\ref{CExp_SoloA_T20_40_g20_cut:fig} and
\ref{CExp_SoloA_T20_40_g20_Gunn_cut:fig}).
To obtain more reliable couplings, an independent estimate of the
vibrational temperature of these ions would be necessary.
It would then be possible to re-compute the electron-phonon parameters by
means of a new fit, including thermal effects.

\section*{\centerline{ACKNOWLEDGEMENTS}}

It is a pleasure to thank P.\ Gattari and E.\ Tosatti for useful
discussions.


\end{document}